\documentclass[11pt]{article}

\usepackage{a4}
\usepackage[utf8]{inputenc}

\usepackage{amssymb}
\usepackage{hyperref}

\usepackage{tikz}
\tikzstyle{every picture}=[level distance = 8mm, baseline=-0.5ex]
\tikzstyle{prop}=[shape=circle,minimum size=6mm, draw=black!80, fill=green!30]

\let\nonu=\nonumber

\begin{document}

\title{Approximate Differential Equations for Renormalization Group Functions 
in Models Free of Vertex Divergencies.}
\author{Marc~P.~Bellon\thanks{UPMC Univ Paris 06, UMR 7589, LPTHE, F-75005,
Paris, France},\thanks{CNRS, UMR 7589, LPTHE, F-75005,
Paris, France}}
\date{}
\maketitle
\begin{abstract}
I introduce an approximation scheme that allows to deduce differential 
equations for the renormalization group $\beta$-function from a 
Schwinger--Dyson equation for the propagator. This approximation is proven to 
give the dominant asymptotic behavior of the perturbative solution.  In the 
supersymmetric Wess--Zumino model and a $\phi^3_6$ scalar model which do not 
have divergent vertex functions, this simple Schwinger--Dyson equation for the 
propagator captures the main quantum corrections.
\end{abstract}

\section{Introduction}\label{int}
New methods for efficient perturbative calculations in Quantum Field Theory are 
badly needed, both for practical and theoretical reasons.  This work explores 
the possibility of approximating Schwinger--Dyson equations as differential 
equations on the renormalization group functions in specific cases.

Four situations will be studied and compared, based on two distinctions: 
the Schwinger--Dyson equation is either linear or quadratic and the theory is 
either a
 four dimensional supersymmetric model or  a six dimensional scalar one. These 
models are characterized by the absence of divergences in vertex functions, at 
all orders for the Wess--Zumino model, only at one loop for the scalar model. 
This allows to avoid Schwinger--Dyson equations for vertex functions which 
would be much more difficult to study.

In the case of linear Schwinger--Dyson equations, Broadhurst and Kreimer 
obtained differential equations for the renormalization group functions 
in~\cite{BrKr99}. A generalization to the case of non-linear Schwinger--Dyson 
equations is the aim of this work. In a preceding work~\cite{BeSc08}, we have 
shown how to obtain numerically high orders of the perturbative solution of a 
nonlinear Schwinger--Dyson equation, elaborating on methods proposed 
in~\cite{KrYe2006}.  These computations showed the singularities of the Borel 
transform of the perturbative series: the singularity on the positive axis is 
weaker than the one predicted by cruder approximations. However the origin of 
the observed asymptotic behavior of the perturbative series was not obvious. In 
this work, I will introduce differential equations whose solutions are good 
approximations for the renormalization group function. Their iterative 
solutions easily show the dominant behavior of the perturbative series at high 
order.

Our previous paper~\cite{BeSc08} was based on two elements, which remain at the 
heart of the present one. A scheme for the exponentiation of the 
renormalization group allows to obtain the full propagator from the 
renormalization group function. The primitive divergence is computed with 
propagators to real powers, yielding a function, the Mellin transform, whose 
Taylor coefficients give the contribution coming from arbitrary logarithmic 
corrections to the propagator. Generically, this Mellin transform is a 
transcendent function, however a rational function can be a suitable 
approximation.
Finally, with a rational Mellin transform, the Schwinger--Dyson equation can be 
converted in a differential equation for the renormalization group function.
Even if these differential equations are the result of an approximation, they 
give the right asymptotic behavior for the perturbative series. It is then easy 
to determine the convergence radius of the Borel transform.

The next section will precise the models and equations we study, and recall the 
results in~\cite{BeSc08} which will be our starting point. Then follows the 
study of the Mellin transform and its singularities. Four approximate 
differential equations are presented with their implications. We conclude with 
the perspectives this work opens, in particular for Schwinger--Dyson equations 
including more primitives. 

\section{Models and methods.}

\subsection{Schwinger--Dyson equations.}

The present study only considers massless models where renormalization is 
limited to the propagator. As in~\cite{BeSc08}, one of these models is the 
supersymmetric Wess--Zumino model. The vertex functions are never divergent in 
this case, as can be proven to all orders in perturbation theory by superspace 
techniques. Another model of interest is the six-dimensional theory of a scalar 
complex field with the interaction:
\begin{equation}
\frac\lambda{3 !}  (\phi^3 + \bar\phi^3)
\end{equation}
The one-loop three-point function is zero: the vertices associated to the 
$\phi^3$ interactions and those associated to its complex conjugate must 
alternate and this is not possible in a three point loop. Furthermore, the 
two-loop contribution to the three-point function is non-planar, so that it 
does not contribute in a large $N$ limit. It is therefore a coherent 
approximation to consider the Schwinger--Dyson equation associated to its 
unique one-loop divergence, which is a propagator correction.

The fundamental object of study of this work is therefore the simplest 
non-linear Schwinger--Dyson equation, which is graphically:
\begin{equation}\label{SDnlin}
\left(
\tikz \node[prop]{} child[grow=east] child[grow=west];
\right)^{-1} = 1 - a \;\;
\begin{tikzpicture}[level distance = 8mm, node distance= 10mm,baseline=(x.base)]
 \node (upnode) [style=prop]{};
 \node (downnode) [below of=upnode,style=prop]{}; 
 \draw (upnode) to[out=180,in=180]  (downnode) 
 	node[name=x,coordinate,midway] {} 
		child[grow=west] ;
\draw (upnode) to[out=0,in=0]  (downnode) 
 	node[coordinate,midway] {} 
		child[grow=east] ;
\end{tikzpicture}
\end{equation}
In this equation, $a$ denotes a suitable equivalent of the fine structure 
constant, which is equal to $\lambda^2$ up to some numerical constant.

The simpler linear Schwinger--Dyson equation, which has been extensively 
studied in~\cite{BrKr99}, will also be considered as a test bed and for 
comparative purposes. It is graphically depicted as 
\begin{equation} \label{SDlin}
\left(
\tikz \node[prop]{} child[grow=east] child[grow=west];
\right)^{-1} = 1 - a \;\;
\begin{tikzpicture}[level distance=14mm]
 \node (upnode) [style=prop]{} child[grow=east] child[grow=west];
\draw (0.8,0) arc(0:180:0.8) ;
\end{tikzpicture}
\end{equation}

In both cases, these equations express a one-particle irreducible two-point 
function in term of an integral over the propagator, which is its inverse. The 
difficulty in providing for solutions of these Schwinger--Dyson equations stem 
from the need to obtain the full propagator, when only its first derivative is 
easily deduced. The gap is bridged by the use of the renormalization group 
combined with a renormalization condition taken at a fixed impulsion.

\subsection{Renormalization group.}

The two-point functions are always considered as a ratio with respect to their 
free counterpart. Adding a renormalization condition at a fixed exterior 
impulsion $p^2=\mu^2$, the two-point functions get an expansion in power of the 
logarithm of the impulsion $L=\log(p^2/\mu^2)$:
\begin{equation}
	G(L) = 1 + \sum_k g_k \frac{L^k}{k!}
\end{equation}
I introduce here a factor $k!$ to take $g_k$ as the 
$k^{\mbox{\scriptsize{th}}}$ derivative of the function, a convention which was 
not in either~\cite{KrYe2006} or~\cite{BeSc08} but proves convenient. Now, the 
renormalization group relates values of the propagator at different impulsions  
and allows to deduce a recursive relation between the $g_k$. In fact, as we 
have seen in~\cite{BeSc08}, similar recursion relations can be found for any 
power, positive or negative, of the propagator. 
 The recursion has two parameters, the power $n$ of the propagator we consider 
and a parameter $b$ which is 2 for the linear Schwinger--Dyson equation and 3 
for the non-linear one. $b$ is the power of the propagator appearing in the 
effective coupling constant:
\begin{equation}\label{recurs}
 g_{k+1} = \gamma (n + b a\partial_a) g_k.
\end{equation}
In this equation, $\gamma$ is the anomalous dimension, the derivative of the 
propagator with respect to the variable $L$. I will note $\gamma_k$ the 
coefficients $g_k$ in the case of the propagator, with the obvious equality 
$\gamma_1=\gamma$.

The proof, detailed in~\cite{BeSc08}, is based on two elements. On the one 
hand, the sums of the diagrams of a given order generate a sub Hopf algebra of 
the renormalization Hopf algebra~\cite{vSu2007} and this property remains true 
for the partial sums including only the diagrams generated by a 
Schwinger--Dyson equation. 

On the other hand, the renormalization group is a one parameter group in the 
group of characters of the renormalization Hopf algebra. This fact is simpler 
to establish in our case than in the original works of Connes and 
Kreimer~\cite{CoKr00}. In the minimal subtraction scheme they used, the 
counterterms are not algebra homomorphism so that it is not trivial that the 
ratio of the counterterms at two different scales gives an algebra 
homomorphism. Here however, renormalization is at a fixed scale, so that the 
counterterm is simply the convolution inverse of the evaluation at the 
renormalization scale. The renormalization group is a simple consequence of the 
definition of the renormalized evaluation, if we know that the renormalized 
evaluation has a well defined limit when the regulator is removed.
\begin{eqnarray*}
\Phi^R_{q^2/p_0^2} &=& (\Phi_{p_0^2}\circ S) \star \Phi_{q^2}\\
&=& (\Phi_{p_0^2}\circ S) \star \Phi_{p^2} \star (\Phi_{p^2}\circ S) \star 
\Phi_{q^2} \\
&=& \Phi^R _{p^2/p_0^2} \star \Phi^R_{q^2/p^2} 
\end{eqnarray*}
In this equation, the convolution inverse is written using the antipode $S$ of 
the renormalization Hopf algebra.

The introduction of these Hopf algebra structures has for me two virtues. It 
allows to give a precise sense to the fact that the very same combinatorial 
identities which make the renormalization program work are at the base of the 
renormalization group. In the case we are interested in, it shows that the 
necessary relations hold when we do not deal with the full perturbative 
development, but only to the part generated by a truncated Schwinger--Dyson 
equation.

\subsection{Evaluation of loop integrals}

In both cases, we simply have to evaluate a single scalar loop integral. This 
is clear for the $\phi^3_6$ case, and was proven in~\cite{BeLoSc07,BeSc08} for 
the supersymmetric one.  Schwinger parameterization allows to obtain the 
following result:
\begin{equation}\label{Mellin}
	F(a,b) = \int  \frac{dk^D }{(k^2)^a \bigl( (p-k)^2 \bigr)^b } = 
\pi^\frac D2 (p^2)^{\frac D2-a-b}\frac{ \Gamma(\frac D2-a)}{\Gamma(a)} \frac{ 
\Gamma(\frac D2-b)}{\Gamma(b)} \frac{ \Gamma(\frac D2-c)}{\Gamma(c)}
\end{equation}
with $c$ defined as $c=D-a-b$. This expression is very symmetric in $a$, $b$ 
and $c$, and this can be explained by the introduction of the two loop vacuum 
diagram with three propagators having the exponents $a$, $b$ and $c$. The 
choice of $c$ makes this diagram scale invariant and hence divergent. This 
infinite result can however be understood as a finite quantity multiplying the 
infinite volume of the dilatation group. If the invariance under the group of 
rotations and dilatations is fixed by choosing one of the impulsions, one 
obtains this finite coefficient. The scale can be fixed in a covariant way with 
$(p^2)^{\frac D2}\delta(p-p_0) $, to obtain a result independent of the 
impulsion which has been fixed. This enhanced symmetry is not very important in 
this simple one-loop case, where an explicit expression for the result is at 
hand, but afford great simplifications for higher loop cases.

At a given order, the propagator takes the form of the free propagator 
multiplied by a polynomial in $L=\log(p^2/\mu^2)$. It can be obtained from the 
action of a differential operator on $(p^2)^{-a}$. More precisely, we write:
\begin{equation}
P(p^2) = \frac 1 {p^2}\Bigl(1 + \sum \gamma_k \frac 
{\log(p^2/\mu^2)^k}{k!}\Bigr) = \left. \Bigl (1 + \sum \gamma_k 
\frac{\partial_x^k}{k!}\Bigr) \frac 1 {(p^2)^{1-x}}\right|_{x=0}
\end{equation}
If we plug this value in the loop integral, we can exchange the derivation and 
the loop integration, so that we end up multiplying the $\gamma_k$ by the 
coefficients of the Taylor expansion around $(0,0)$ of $F(1-x,1-y)$.

The object of interest is therefore the Taylor expansion of $F(1-x,1-y)$. The 
function $F$ has poles whenever one of the $\Gamma$ function in the numerator 
has a negative integer as argument, i.e., for the parameters equal to $\frac 
D2$ or greater integers. The residues are simple polynomials. When $a$ or $b$ 
is a great enough integer, it corresponds to infrared divergences. For $c$ 
instead, the pole comes from a ultraviolet divergence, but the two are linked 
by conformal invariance. This link between poles of the Mellin transform and 
divergences of the diagram gives clues for the computation of the residues even 
at higher loop order, when the full Mellin transform has no closed form 
expression. This should allow to control the effect of higher loop corrections 
to the Schwinger--Dyson equation and will be the subject of future work.

\section{The differential equations}

\subsection{Linear Schwinger--Dyson equations}

In the cases with linear Schwinger--Dyson equations,  differential equations 
for the anomalous dimension were presented in~\cite{BrKr99}. However the 
process through which these equations were encountered, involving explorations 
and computer assisted transformations of a system of partial differential 
equations, was not enlightening. A first explanation was presented in this same 
work,  through the notion of propagator--coupling duality: this gives the 
equations we derived from the renormalization group. 

The four dimensional case is really simple, without any possible variation, and 
is left as an exercise for the interested reader. In the $\phi_6^3$ case, 
equation~(\ref{Mellin}) is taken with $b=1$, $a=1-x$ and $D=6$. The Mellin 
transform becomes a simple rational function. The Schwinger--Dyson equation 
takes the following form:
\begin{equation}\label{linC}
	\Pi(L)= 1 - a\bigl( \sum \gamma_k \frac{\partial_x^k}{k!}\bigr)[ 
e^{Lx}-1] \frac1{x(1+x)(2+x)(3+x)}
\end{equation}
Conformal invariance corresponds to the exchange of $a$ and $c$ and implies a 
symmetry for $x$ transformed to $-3-x$.
The application of a suitable  differential operator with respect to $L$ allows 
to reduce the right hand side to a constant, and the different terms of the 
left hand side can be obtained from $\gamma$ through the use of some version of 
equation~(\ref{recurs}). This derivation has been presented in Yeats' 
thesis~\cite{Ye08}. 

This case will be used to test the approximation scheme I propose: let us 
forget the poles most removed from the origin and compensate with the first few 
terms of the Taylor expansion. One obtains a simpler differential equation. 
Instead of applying the full differential operator 
$\partial_L(1+\partial_L)(2+\partial_L)(3+\partial_L)$, we drop the last 
factors and truncate the Taylor series for $1/(2+x)(3+x)$ to obtain a 
differential expression on $\gamma$ of similar order. Instead of 
\begin{equation}
\label{LSD1}
\bigl (3+\gamma(2a\partial_a-1)\bigr) \bigl (2+\gamma(2a\partial_a-1)\bigr) 
\bigl (1+\gamma(2a\partial_a-1)\bigr) \gamma = a,
\end{equation}
one obtains:
\begin{equation}
\label{LSD3}
	\gamma + \gamma(2a\partial_a-1)\gamma = \frac a 6 \Bigl( 1 - \frac 5 6 
\gamma + \frac {19}{36} 
		\gamma(2 a \partial_a +1) \gamma \Bigr)
\end{equation}
Let us remark that the terms quadratic in $\gamma$ on the left and right hand 
sides of this equation differ, since in one case, one is dealing with the 
higher derivatives of the 1PI two-point function and in the other one, with 
those of its inverse, the full propagator. A comparison of numerical solutions 
of the two equations show that the simplified equation~(\ref{LSD3}) fairs 
remarkably well: the two solutions are visually indiscernible up to $a=2$, the 
relative error is around half of a percent at $a=1$. The behaviors however 
diverge for larger $a$. When the exact equation~(\ref{LSD1}) has a regular 
solution for large $a$, behaving as $a^{1/2}$ up to logarithmic corrections, 
the solution of the approximate equation~(\ref{LSD3}) hits a singularity around 
$a = 11.3684$.

The perturbative solutions can also be compared. As in~\cite{BrKr99}, the 
coefficients of degree $n$ are multiplied by $(-1)^n 6^{2n-1}$ to get natural 
numbers, beginning with 1, 11, 376. These first coefficients are equal, since 
the approximation is exact up to this order. The coefficients for the 
approximate solution are comparable to the one of the full solution, with a 
ratio decreasing slowly as $n^{-\beta}$, with $\beta$ around $0.22$. This value 
stem from the computation of 200 coefficients. One could also study the 
intermediate situation where only the $(3+\partial_L)$ factor is dropped, but 
this is of limited interest.

What I wanted to show is that with the action of a suitable differential 
operator such that the Taylor coefficients  of the Mellin transform become 
rapidly small, truncating the Taylor expansion to a few terms nonetheless 
produces sensible results. Furthermore, an equation as~(\ref{LSD3}) allows to 
compute easily the ratio of two successive terms in the perturbative expansion, 
which is $n/3$. This would be less clear from the full equation~(\ref{LSD1}).

\subsection{Non-linear Schwinger--Dyson for $\phi_6^3$}

In the case of a nonlinear Schwinger--Dyson equation for the six-dimensional 
model, the relevant Mellin transform is:
\begin{equation}
\label{mell6}
e^{L(1+x+y)} \frac{\Gamma(2+x)}{\Gamma(1-x)}  \frac{\Gamma(2+y)}{\Gamma(1-y)}  
\frac{\Gamma(-1-x-y)}{\Gamma(4+x+y)}
\end{equation}
The poles associated to the factors $\Gamma(2+x)$ or $\Gamma(2+y)$ are farther 
from the origin and give contributions to the Taylor expansion which decrease 
at least as $2^{-n}$ at order $n$. The dominant contributions for the high 
order of the Taylor expansion come from the poles of $\Gamma(-1-x-y)$. As in 
the preceding case, we can differentiate with respect to $L$ in order to cancel 
the poles near the origin. More precisely, we need to multiply the Mellin 
transform by $(-1-x-y)(-x-y)(1-x-y)$ which corresponds to the action of 
$\partial_L - \partial_L^3$. Using the relevant renormalization group equation 
for the higher derivatives of $\gamma$ and a suitable truncation of the Taylor 
expansion one obtains:
\begin{eqnarray}
	\gamma -\gamma ( 3 a \partial_a -1 ) \gamma (3 a \partial_a -1) \gamma 
&=& \nonumber \\
		\frac a6 \biggl(1- \frac {11}{3} \gamma &+& \frac {a}{18} 
\bigl( 49 \gamma(3a\partial_a +1) \gamma + 67 \gamma^2\bigr) \biggr)
\end{eqnarray}
The consequences of this equation will not be detailed here, since one lacks 
suitable comparison points. However, it is easy to see that for large orders in 
$a$, the dominant term in the expansion of $\gamma$ comes from the cubic in 
$\gamma$ term on the left hand side. Associated to the lowest order of $\gamma$ 
which is $a/6$, this proves that the ratio of successive terms is 
asymptotically $-n/2$. This fixes the convergence radius of the Borel transform 
of the perturbative series and indicates that the main singularity is on the 
negative axis.

\subsection{The Wess--Zumino model, simple equation.}

The $\gamma$ function for this model was studied in~\cite{BeSc08} and a number 
of observations could be made on the behavior of the resulting series. The 
approximations we propose here will be checked against the detailed 
computations we made, and reciprocally, the approximations allow to prove the 
observed properties of the series. 

Our starting point is the Mellin transform obtained in~\cite{BeSc08}:
\begin{equation}\label{depart}
- (e^{L(x+y)}-1) \frac{\Gamma(1+x)\Gamma(1+y) 
\Gamma(-x-y)}{\Gamma(1-x)\Gamma(1-y) \Gamma(2+x+y)}
\end{equation}
This case presents a new problem. This Mellin transform presents a pole for 
$x+y=1$, but the dominant ones are the poles for $x=-1$ and symmetrically 
$y=-1$. This poles cannot be cancelled by derivations with respect to $L$, as 
the ones depending only on $x+y$. The residue of the pole for $x=-1$ is $1$ for 
the above expression, but becomes $1-y$ or $1+xy$ due to the derivation with 
respect to $L$ necessary to cancel the divergence for $x=y=0$.

In a first step, let us consider the contribution coming only from 
$(1+x)^{-1}$. The application of the differential operator $\sum \gamma_k 
\partial_x^k/k!$ gives the sum of the $(-1)^k\gamma_k$.
Since $\gamma_{k+1}$ is deduced from $\gamma_k$ by the application of the 
operator $\gamma(3a\partial_a +1)$, we obtain the formal series:
\begin{equation}
\sum_k(-1)^k \gamma_k = \sum_k (-1)^k \bigl[ \gamma(3a\partial_a+1)\bigr]^k 1 = 
\frac 1 {1+\gamma(3a\partial_a +1 )} 1.
\end{equation}
The exact definition of the inverse does not matter, since we will be multiply 
by the operator to cancel it.  However, since the sum of the $\gamma_k$ is 
multiplied by $a$, the operator must be permuted with the operation of 
multiplying by $a$. We therefore multiply both sides of the equation by the 
operator $1+\gamma(3a\partial_a - 2) $. Putting it all together, and adding 
polynomial contributions to have a result exact up to the third order, we 
obtain successively:
\begin{eqnarray}
 \gamma &=& 2 a \gamma ^2 -a + 2 a \frac1 { 1 + \gamma (3 a \partial_a +1)} 1 
\label{WZ1}\\
&& \bigl [1 + \gamma(3 a \partial_a -2) \bigr] \bigl(\gamma + a - 2 a \gamma^2 
\bigr) = 2a \\[1.7mm]
\gamma &=& a - a \gamma - \gamma (3a\partial_a -2) \gamma + 2 a  \gamma^2 +2 a 
(2 a \partial_a + 1 ) \gamma^3 \label{WZs}
\end{eqnarray}

With this formula, it is easy to obtain the asymptotic growth of the 
coefficients in the development of $\gamma = \sum_n (-1)^{n-1}c_n a^n$. The 
term $\gamma (3a\partial_a -2) \gamma$ gives the ratio of successive terms 
proportional to $3n$. The next term for this ratio is easy to compute, since 
the term cubic in $\gamma$ and $a \gamma^2$ do not contribute at this level of 
precision. One obtains:
\begin{equation}
c_{n+1} \simeq (3n+2) c_n \label{asymp}
\end{equation}
This is exactly the result obtained experimentally from the calculation 
in~\cite{BeSc08}. The fact that this most simple approximation has this ratio 
asymptotically exact up to the constant terms is at first a surprise. It 
nevertheless has the interesting consequence that the ratio of these 
approximate coefficients and the exact ones will reach a finite limit.  Indeed, 
a product of terms which behaves asymptotically as $1+ O(1/n^2)$ is convergent. 
The comparison of the coefficients obtained from the iterative solution of 
equation~(\ref{WZs}) with the more precise results obtained in~\cite{BeSc08} 
indeed shows that their ratio, which starts at 1 for the first terms, has a 
limit which can be estimated to be $0.942617$ after hitting a maximal deviation 
at $0.9250$. 

In the six-dimensional theory, it was more difficult to obtain this level of 
precision. This can be understood from the relative sizes of the terms at a 
given order in $a$. 
The rapid growth of the coefficients of $\gamma$ makes the term of order $n$ of 
a product dominated by the term with one of the factors of the highest possible 
order.  The recursive definition of $\gamma_k$ than shows that the coefficients 
of $a^n$ in all the $\gamma_k$  are of comparable size when $n$ is larger than 
$k$. In a product $\gamma_1 \gamma_k$, the dominant contribution will come from 
the term of degree $n-1$ in $\gamma_k$, which is a factor of order $n$ smaller, 
and products involving $\gamma_2$ will make still smaller contributions. 

The terms which are linear in the $\gamma_k$ are therefore asymptotically 
dominant. In the Wess--Zumino model, the pole at $x=-1$ gives the exact 
contribution for $y=0$ and therefore for this dominant terms, whereas in the 
$\phi_6^3$ model, two further poles are necessary to obtain the full 
contribution. 

We should nevertheless expect that the terms coming from $xy/(1-x-y)$ and 
$xy/(1+x)$ contribute finite terms to equation~(\ref{asymp}), since they 
represent $n$ terms of size $1/n$. However, $xy/(1-x-y)$ has all its Taylor 
coefficients positive so that there are cancellations due to the alternating 
signs of the $\gamma_k$ for a given order. The cases of $xy/(1+x)$ and 
$xy/(1+y)$ are more subtle. The corresponding contributions can be written as:
\begin{equation}
 a \; \gamma \frac 1 {1+\gamma (3 a \partial_a+1) } \gamma 
\end{equation}
 When they are multiplied by the operator $1+\gamma(3a \partial_a-2)$, the 
dominant infinite series is again cancelled for this term also, and we remain 
with
 \begin{equation}
3 a( \gamma a \partial_a \gamma)\frac 1 {1+\gamma (3 a \partial_a+1) } \gamma 
\end{equation}
The infinite series of terms is therefore multiplied by a term of order $a^3$, 
so that the coefficients are individually proportional to $c_n/n^2$, and their 
sum cannot contribute a finite term. We have thus shown that the differential 
equation~(\ref{WZs}) allows to predict the ratio of the successive terms of the 
series for $\gamma$ up to vanishing terms.

\subsection{The Wess--Zumino model: higher precision equation.}
In order to reach a higher precision on the asymptotic behavior, it is 
necessary to take into account the pole for $x+y=1$ and the full residue of the 
poles at $x$ or $y=-1$. Canceling the pole for $x+y=1$ can be achieved simply 
by taking derivatives with respect to $L$, the logarithm of the impulsion, to 
multiply the Mellin transform by $(x+y)(1-x-y)$. This however add to the 
complexity of the residue of the poles at $x$ and $y=-1$.
The Mellin transform now reads:
\begin{eqnarray}\label{mellinWZf}
	\frac{ \Gamma(2-x-y) \Gamma(1+x)\Gamma(1+y) }{ \Gamma(2+x+y) 
\Gamma(1-x)\Gamma(1-y) } &=&
\nonu\\
	\frac{(1-x)(2-x)}{1+y}\!& +&\! \frac{(1-y)(2-y)}{1+x} -3 + 3(x+y)\\ 
-(x+y)^2
	\! &+&\! 2\bigl(\zeta(3)-1\bigr)xy(x+y) +\cdots\nonu
\end{eqnarray}
In our preceding work, we remarked that expressing the residues in terms of the 
product $xy$ allowed for a simpler polynomial part, but in the present case, it 
is better to have the different summands of the residue give similar terms. The 
corresponding equation for the $\gamma$ function, dropping the term 
proportional to $\zeta(3)-1$, is:
\begin{eqnarray}
\gamma - \gamma( 3 \nabla -1 ) \gamma &=& -3a + 6a\,\gamma - 2 
a\,\gamma(3\nabla +2) \gamma \nonu\\
&+&a \,\bigl( 4 - 6 \gamma  + 2 \gamma (3\nabla + 1 ) \gamma \bigr) \frac 1 { 1 
+\gamma(3\nabla +1 )} 1
\end{eqnarray}
The operator $\nabla$ has been introduced as a short hand for $a\partial_a$ to 
keep down the size of the equation.  The formal inverse can be removed by 
putting everything else on the other side:
\begin{eqnarray}\label{WZf2}
\frac {\gamma + 3 a - \gamma(3\nabla-1)\gamma- 6a \, \gamma + 2 a 
\,\gamma(3\nabla+2)\gamma}
{4 - 6 \gamma  + 2 \gamma (3\nabla + 1 ) \gamma} &=& \frac 1 { 1 + 
\gamma(3\nabla -2)} a
\end{eqnarray}
In this form, the differential equation obtained by applying 
$1+\gamma(3\nabla-2)$ to both sides looks rather daunting. The presence of a 
quotient reintroduces the necessity of series inversion that we avoided by a 
clever use of the renormalization group equations. Otherwise, the expansion of 
the derivative of the quotient, followed by a multiplication by the square of 
the denominator in~(\ref{WZf2}) gives a polynomial equation, but with numerous 
terms.

Let us remark that in any case, the derivatives with respect to $a$ get 
multiplied by $ a \gamma$: the total number of possible terms of a given degree 
in $\gamma$ is therefore limited. It is possible that the combinatorial methods 
of the operad of algebras with derivation introduced by Jean-Louis 
Loday~\cite{Lo2009} is useful to stitch together similar terms. The complexity 
of the obtained equation raises the question whether a systematic improvement 
of such approximations by the addition of the contribution of other poles of 
the Mellin transform is practical.

The factor $a\gamma$ coming with each derivatives has a double consequence. On 
one side, it ensures that perturbatively, higher derivative terms are 
subdominant, but this also makes the differential equation highly singular in 
the vicinity of $a=0$: proving the non-perturbative existence of the solution 
is not straightforward.

\section{Conclusion.}

In the present paper, I have shown how to deduce from Schwinger--Dyson 
equations simple differential equations for the renormalization group 
functions. They readily give the asymptotic behavior of the perturbative series 
and in particular the convergence radius of the Borel transform. Through the 
inclusion of the contributions of more poles of the Mellin transform, it should 
be possible to obtain systematic improvements of the solution.

Differential equations for renormalization group functions had been proposed in 
recent years. In her thesis~\cite{Ye08}, Karen Yeats proposed a way to 
linearize the nonlinear Schwinger--Dyson equations and obtain simple 
differential equations for the renormalization group functions. The proposition 
has been applied both to QED~\cite{BaKrUmYe08} and to QCD~\cite{BaKrUmYe09}. 
However, the transformed Schwinger--Dyson equation is indeed linear, but with 
an infinite number of terms and an unknown function appears in the differential 
equation, with a very complex recursive definition. It is therefore not clear 
if in a perturbative solution, the contribution from the non-linear 
differential term dominates the contribution of the unknown function.

Up to now, we only considered simple Schwinger--Dyson equations, with a 
one-loop correction to the propagator. However, the full Schwinger--Dyson 
equation includes higher order terms, and we would like to know how these 
additional contributions modify the properties of the renormalization group 
functions. The difficulty a priori with such terms is the great number of 
propagators, each coming with its own variable, and therefore the rapid growth 
of the number of terms with a given number of derivatives of the Mellin 
transform. However the leading contributions in the Taylor expansion of the 
multivariable Mellin transforms correspond to its poles, which can be related 
to the divergences of the diagram. Indeed, with every propagator coming with a 
variable exponent, all subdiagrams become divergent for some choice of the 
exponents. A rôle should be find for the core Hopf algebra introduced 
in~\cite{Kreimer:2009jt, Kreimer:2009iy} to organize these divergences. The 
poles have a simple structure, because they only depend on the sum of the 
Mellin variables of a given subgraph. The highly nonlinear character of such 
Schwinger--Dyson equations should not be a hindrance to their successful use. 
In particular, at least in a large $N$ limit where the number of primitive 
divergences does not grow too fast, it could be possible to show that these 
additional terms do not change the leading asymptotic behavior of the 
perturbative series.

Another desirable extension is to deal with vertex renormalization. However the 
vertices depend a priori on different energy scales and the full vertex is  not 
entirely defined by its renormalization group dependence. There are also 
overlapping divergences, which mean that it is not possible to simply replace 
the sum of a vertex and its counterterm by a renormalized vertex. We must also 
choose the renormalization point for the vertex. In QED, the Ward identities 
are simpler for the vertex with a zero impulse photon, but this is not a 
suitable choice in a massless theory. 

Whatever the successes we encounter in these improvements, this work has 
already delivered. It has shown how to combine Schwinger--Dyson equations and 
renormalization group to control the asymptotic behavior of the perturbative 
series for  exactly renormalizable quantum field theories which are not so 
artificial than the ones studied in~\cite{BrKr99,BeLoSc07}. This is a 
remarkable result, since the simple recursions developed here subsume huge 
number of individual graphs with their hierarchy of counterterms, with 
important cancellations between contributions of differing signs.
 
 \bigbreak
\noindent{\bf Acknowledgments:} I wish to express my special thanks to Olivier 
Babelon who suggested to look for global properties of the Mellin transform in 
order to understand the asymptotic properties of its local expansion which 
puzzled us in our preceding work. The presentation of our previous work in 
seminars helped me to clarify the concepts: thanks to all who invited me and 
specially to the organizers of a workshop in Cargèse.

\end{document}